\input harvmac
\overfullrule=0pt
\def\sqr#1#2{{\vbox{\hrule height.#2pt\hbox{\vrule width
.#2pt height#1pt \kern#1pt\vrule width.#2pt}\hrule height.#2pt}}}
\def\Box{\mathchoice\sqr64\sqr64\sqr{4.2}3\sqr33}

\def\half{{\textstyle{1\over 2}}}

\Title{ \vbox{\baselineskip12pt
\hbox{hep-th/0210030}}}
%\hbox{IASSNS-HEP-99/66}
%\hbox{CAL}}}
{\vbox{\centerline{Noncommutativity in
a Time-Dependent Background}}}
\bigskip
\centerline{Louise Dolan\foot{dolan@physics.unc.edu}}
\smallskip
\centerline{\it Department of Physics}
\centerline{\it
University of North Carolina, Chapel Hill, NC 27599-3255}
\bigskip
\smallskip
\centerline{Chiara R. Nappi\foot{cnappi@princeton.edu}}
\smallskip
\centerline{\it Department of Physics, Jadwin Hall}
\centerline{\it Princeton University, Princeton, NJ 08544-0708}

\bigskip

\noindent
We compute a time-dependent noncommutativity parameter
in a model with a time-dependent background, a space-time metric of the 
plane wave type supported by a Neveu-Schwarz two-form potential.
This model is an open string version of the WZW model based
on a non-semi-simple group previously studied by Nappi and Witten.    
The background we study is not conformally invariant.
We consider a light-cone action for the sigma-model,
compute the worldsheet propagator, and use it to exemplify a
derivation of a time-dependent noncommutativity parameter. 
\Date{}
%references

\nref\witten{E. Witten, ``Noncommutative Geometry and String Field Theory,''
Nucl. Phys. {\bf B268} (1986) 253.} 

\nref\connes{A. Connes, M.R. Douglas, and A. Schwarz, 
``Noncommutative Geometry and Matrix Theory: Compactification on Tori,'' 
JHEP 9802 (1998) 003, hep-th/9711162.}

\nref\nek{M. Douglas and N. Nekrasov, ``Noncommutative Field Theory'',
Rev. Mod. Phys. {\bf 73} (2001) 977, hep-th/0106048.}

\nref\sw{N. Seiberg and E. Witten, ``String Theory and Noncommutative
Geometry'', JHEP {\bf 9909} (1999) 032,  hep-th/9908142.}

\nref\acny{A. Abouelsaood, C. Callan, C. Nappi, and S. Yost,
``Open Strings in Background Gauge Fields''
Nuclear Physics {\bf B280} (1987) 599.}

\nref\hs{G. Horowitz and A. Stief, ``Strings in Strong Gravitational
Fields'', Physical Review {\bf D42} (1990) 1950.}

\nref\sen{A. Sen, ``Time Evolution in Open String Theory'',
hep-th/0207105.}

\nref\perkr{S. Hemming, E. Keski-Vakkuri, P. Kraus,
``Strings in the Extended BTZ Space-Time'', hep-th/0208003.}

\nref\coscor{L. Cornalba and M. Costa, ``A New Cosmological Scenario
in String Theory'', Phys. Rev. {\bf D66} (2002) 066001, hep-th/0203031.} 

\nref\ffjs{J. Figueroa-O'Farrill and Joan Simon, ``Generalised Supersymmetric
Fluxbranes'', JHEP 0112 (2001) 011, hep-th/0110170.}

\nref\shash{A. Hashimoto, S. Sethi,
``Holography and String Dynamics in Time Dependent Backgrounds'',
hep-th/0208126.}

\nref\ap{M. Alishahiha and S. Parvizi, ``Branes in Time-Dependent
Backgrounds and AdS/CFT Correspondence'', hep-th/0208187.}

\nref\sm{H. Liu, G. Moore, and N. Seiberg,
``Strings in Time Dependent Orbifolds'',
hep-th/0204168, hep-th/0206182.}

\nref\hp{G. Horowitz and J. Polchinksi,
``Instability of Space-Like and Null Orbifold Singularities'',
hep-th/0206228.}

\nref\gpz{E. Gimon, L. Pando Zayas and J. Sonnenschein,
``Penrose Limits and RG Flows'', hep-th/0206033.}

\nref\lr{H. Lewis and W. Riesenfeld,
``An Exact Quantum Theory of the Time-Dependent Harmonic
Oscillator and of a Charged Particle in a Time-Dependent Electromagnetic
Field'', Jour. Math. Phys. {\bf 10} (1969) 1458.}

\nref\kl{D. Khandekar and S. Lawande, ``Exact Propagator for a
Time-Dependent Harmonic Oscillator with and without a Singular
Perturbation'', Jour. Math. Phys. {\bf 16} (1975) 384.}

\nref\dhh{C. Duval, Z. Horvath, and P. Horvathy, ``Vanishing of the
Conformal Anomaly for Strings in a Gravitational Wave'', Phys. Lett.
{\bf B313} (1993) 10.}

\nref\dhht{C. Duval, Z. Horvath, and P. Horvathy, ``Strings in Plane-fronted
Gravitational Waves'', Mod. Phys. Lett. {\bf A8} (1993) 39.}

\nref\nw{C. Nappi and E. Witten, ``A WZW Model Based on a Non-semi-simple
Group'', Phys. Rev. Lett. {\bf 71} (1993) 3751;  hep-th/9310112.}

\nref\dn{L. Dolan and C. Nappi, ``A Scaling Limit With
Many Noncommutativity Parameters, Phys. Lett. {\bf B504} (2001) 329;
hep-th/0009225.}

\nref\fhhp{P. Forgacs, P. Horvathy, Z. Horvath, and L. Palla,
``The Nappi-Witten string in the light-cone gauge'', hep-th/9503222.}

\nref\tr{J.G. Russo and A.A. Tseytlin, ``Constant Magnetic Field in 
Closed String Theory: An Exactly Solvable Model'', Nucl. Phys. {\bf B448}
(1995) 293, hep-th/9411099.}

\nref\hasht{A. Hashimoto and K. Thomas, ``Dualities, Twists, and Gauge
Theories with Non-Constant Non-Commutativity'', hep-th/0410123.}

\nref\stt{S. Stanciu and A. Tseytlin, ``D-branes in Curved Spacetime:
Nappi-Witten Background,'' JHEP {\bf 9806}, 010 (1998)
hep-th/9805006.}

%\nref\th{L. Thorlacius, ``Born-Infeld String as a Boundary Conformal
%Field Theory'', Phys. Rev. Lett. {\bf 80} (1998) 1588; hep-th/9710181.}

\nref\sch{V. Schomerus, ``D-Branes and Deformation Quantization,'' JHEP
{\bf 9906} (1999) 030; hep-th/9903205.}

%paper
\newsec{Introduction}

Noncommutativity in string theory is a very interesting topic, 
as it may have important implications for the structure of spacetime.
Noncommutativity has emerged in the context of open strings,
starting from the treatment of open string field theory in {\refs{\witten}}.
More recently, it has reappeared in the context of Matrix theory compactified 
on a torus {\refs{\connes,\nek}}, and in the 
low energy description of strings in an electromagnetic background
{\refs{\sw, \acny}}. 

It is interesting to find other models in which noncommutativity emerges.
In most of the examples currently known, 
the noncommutativity parameter is constant. 
An obvious task is to look for time-dependent noncommutativity parameters, 
especially given the recent interest in strings on time-dependent backgrounds 
{\refs{\hs-\dhht}}.

In this paper we study an 
open string model, whose target space has a plane wave metric
supported by a time-dependent Neveu-Schwarz two-form potential.
This background was studied by Nappi and Witten {\refs{\nw}} for closed
strings. Here we are looking at an open string version, and by computing the
worldsheet propagator we can derive a time-dependent noncommutativity
parameter. It is important that the background is of the Neveu-Schwarz type:
plane waves with Ramond fields remain commutative 
as the Ramond background amounts to the addition of a mass term to the
action in light-cone gauge. In our case, for large values of
the time parameter, our model reduces to
a neutral string in a constant background $B$ field {\refs{\sw,\dn}}, 
hence it is a good candidate for space-time noncommutativity.

In sect.2, we consider a light-cone action of the model.
The mode expansion of a  
closed string version of this model has been explicitly exhibited in
{\refs{\fhhp, \tr}}. 
We compute the open mode expansion as a power series
in a suitable parameter $\mu$. This expansion is adequate to 
show noncommutativity. In sect.3
the worldsheet propagator is derived on the disk. In sect.4 we
evaluate the propagator on the boundaries and
compute a time-dependent noncommutativity parameter.
The techniques used in this calculation are similar to those of 
{\refs{\dn}} which analyzes strings in a 
$U(1) \times  U(1)$ background.

\newsec{A Time-Dependent Background}

The Polyakov action coupling a string to a general metric and
background Neveu-Schwarz field is
\eqn\poly{\eqalign{S =
%{-{1\over{4\pi\alpha'}}}
\int_\Sigma\, d\tau d\sigma
\,[\,{\sqrt{-\gamma}}\gamma^{\alpha\beta}\, G_{MN} \partial_\alpha
X^M \partial_\beta X^M + B_{MN} \epsilon^{\alpha\beta} \partial_\alpha X^M
\partial_\beta X^N \,]}}
where we choose the string worldsheet $\Sigma$ with Lorentz signature,
and have rescaled the scalar worldsheet fields by $(2\sqrt{\pi\alpha'})^{-1}$
so that the $X^M$ are dimensionless.
We consider the time-dependent background provided by
the Nappi Witten WZW model based on a non-semi-simple group, 
and adopt the same notation as in{\refs{\nw}},
with $X^M = (a_1,a_2,u,v)$, 
and $u$ being identified with the time in the target space.
\eqn\bmet{G_{MN} =\pmatrix{1&0&{a_2\over 2}&0\cr
0&1&{\hskip-3pt-{a_1\over 2}}&0\cr
{a_2\over 2}&{\hskip-3pt-{a_1\over 2}}&b&1\cr
0&0&1&0\,\cr}\,,
\qquad B_{MN} =\pmatrix{0&u&0&0\cr
-u&0&0&0\cr
0&0&0&0\cr 0&0&0&0\cr}\,.} The Lorentz signature target space metric
$G_{MN}$ can be recognized as a plane wave metric {\refs{\nw}}.
The time-dependence is
the $u$-dependence of $B_{12}$.
Nappi and Witten checked that this model is exactly conformally
invariant ({\it i.e.} to all orders in $\alpha'$) by showing  
the one-loop $\beta$ function equations for the closed string backgrounds
were satisfied,
and then proving there were no higher order graphs.

In this paper, since we are interested in noncommutativity,
we consider open string boundary conditions.
Our case is not conformally invariant, and
the background \bmet\ satisfies the
the Born-Infeld field equations\footnote*{We thank A. Hashimoto and K. Thomas
{\refs{\hasht}} for pointing out an error in Eq.(2.3) 
in a previous version of this paper. A conformally invariant version of 
\bmet\ is studied in {\refs{\stt}}, 
but its noncommutativity parameter although
non-constant, is not time-dependent.} only for $N\ne u$  

\eqn\bi{(D_M F_{NL}) ( 1 - F^2)^{-1 \,LM} = 0}
where $ ( 1 - F^2 )^{-1\,LM} = ( 1 + F )^{-1\,LP}\, G_{PN}\,
(1 - F)^{-1\,NM}$ and
$( 1 - F)_{MN} \equiv G_{MN} - 2\pi\alpha' F_{MN}\,.$
In our case $F_{MN} = B_{MN}$. For \bmet\
the nonvanishing components of the Ricci tensor and
affine connections are $R_{uu} = -\half$,
$ \Gamma^i_{uj} = \half\epsilon^i_{\,j}$, $\Gamma^v_{ui} = -{a^i\over 4}\,.$
It follows that
${(D_M F_{NL}) ( 1 - F^2)^{-1 \,LM}
= \epsilon_{ij}  ( 1 - F^2)^{-1 \,ju} = 0\,,}$ for $N\ne u$.
But 
\eqn\nbi{(D_M F_{uL}) ( 1 - F^2)^{-1 \,LM} = -{u\over{1 + u^2}}
\,.} 

As in {\refs{\nw}}, the
sigma model action is \poly\ :
\eqn\polynw{\eqalign{S&=
%{-{1\over{4\pi\alpha'}}}
\int_\Sigma\, d\tau d\sigma
\,[\,{\sqrt{-\gamma}}\gamma^{\alpha\beta}\,(\partial_\alpha a^i\partial_\beta
a^i + 2 \partial_\alpha u\partial_\beta v +
b \partial_\alpha u\partial_\beta u  + \epsilon_{ij} \partial_\alpha u 
\partial_\beta a^i a^j)
+\epsilon^{\alpha\beta} \epsilon_{ij} u
\partial_\alpha a^i \partial_\beta a^j\,]
\,.\cr}}

Although this action has a cubic interaction, 
if one treats it as a closed string theory, it is possible to find an 
exact mode expansion in the light-cone gauge {\refs{\fhhp, \tr}}. 
However, in considering it as an open string theory, 
one has different boundary conditions 
which make the solution more complicated.
Consequently,
we will solve the theory only via
a power series expansion. For simplicity, we work to lowest order in $\mu$, 
where $\mu$ is a dimensionless constant,
as this is sufficient to see noncommutativity. 
It is quite possible that another 
version of this model, differing from \polynw\ via boundary terms, 
would lead to an exact mode expansion.

Although our background is not conformally invariant, 
we will consider a light-cone version of the sigma model 
in order to study open string propagators in a B-field 
with linear time dependence. We let 
\eqn\lcone{u = \mu\tau\,,} and write
\eqn\lcact{S_{lc} = \int_\Sigma\, d\tau d\sigma
\,[\,\eta^{\alpha\beta}\, \partial_\alpha a^i\partial_\beta
a^i - 2 \mu \partial_\tau v - b \mu^2 
- \mu \epsilon_{ij} \partial_\tau a^i a^j + 2 \,\epsilon_{ij} \,\mu \tau\,
\partial_\tau a^i \partial_\sigma a^j ]\,.} 
%To implement light-cone gauge, we find the Virasoro constraints
%from varying \polynw\ with respect to $\gamma_{\alpha\beta}$.
%In orthonormal gauge $\gamma_{\alpha\beta} = \eta_{\alpha\beta}$,
%they are given by 
%\eqn\vircon{\partial_\alpha X^M \partial_\beta X^M G_{MN}
%-\half \eta_{\alpha\beta} \eta^{\gamma\delta} \partial_\gamma X^M
%\partial_\delta X^N G_{MN} = 0}
%for the background \bmet.
Here $\Sigma$ is the string worldsheet with
Minkowski metric $\eta^{\alpha\beta}$ with non-vanishing components 
$\eta^{\tau\tau} = -1, \eta^{\sigma\sigma} = 1$. We will use
%$\Box\equiv -\partial_\tau^2 + \partial_\sigma^2$.
%In orthonormal gauge, \poly\ becomes
%\eqn\polyog{\eqalign{S&=
%\int_\Sigma\, d\tau d\sigma
%\,[\,\eta^{\alpha\beta}\,( \partial_\alpha a^i\partial_\beta
%a^i + 2 \partial_\alpha u\partial_\beta v +
%b \partial_\alpha u\partial_\beta u + \epsilon_{ij} \partial_\alpha u 
%\partial_\beta a^i a^j)
%+\epsilon^{\alpha\beta} \epsilon_{ij} u
%\partial_\alpha a^i \partial_\beta a^j\,]
%\cr}}
$\epsilon^{\tau\sigma} = 1$, and for the open string
$-\infty\le\tau\le\infty$, $\, 0\le\sigma\le\pi$.
%The equations of motion and Neumann boundary conditions
%obtained by extremizing \polyog\ with respect
%to $X^M(\sigma,\tau)$ are
%\eqn\eomorth{\eqalign{\Box a^i +\half \epsilon_{ij} a^j \Box u
%+ \epsilon_{ij} ( \eta^{\alpha\beta} + \epsilon^{\alpha\beta} )
%\partial_\alpha u \partial_\beta a^j &= 0\cr
%\partial_\sigma a_i + \half \partial_\sigma u \epsilon_{ij} a^j
%- \epsilon_{ij} u \partial_\tau a^j \,|_{\sigma=0,\pi} &= 0\cr
%\Box v + b \Box u +\half \epsilon_{ij} a^j \Box a^i - \half \epsilon_{ij}
%\epsilon^{\alpha\beta} \partial_\alpha a^i \partial_\beta a^j &= 0\cr
%\partial_\sigma v + b \partial_\sigma u +\half \epsilon_{ij} a^j
%\partial_\sigma a^i \, |_{\sigma=0,\pi} &= 0\cr
%\Box u = 0\,,\qquad \partial_\sigma u  \, |_{\sigma=0,\pi} &= 0\,.\cr}}

%As in flat target space, here we can use the residual worldsheet gauge
%invariance to choose the light-cone gauge condition: $u = \mu\tau$, for
%$\mu$ is a dimensionless constant.
%In this gauge we can solve the constraints \vircon\ for the dependent
%variable $v$:
%\eqn\virconlc{\eqalign {\mu\partial_\tau v &= -\half \partial_\tau a^i
%\partial_\tau a^i  -\half \partial_\sigma a^i \partial_\sigma a^i
%- {b\over 2} \mu^2 -\half \mu \epsilon_{ij} \partial_\tau a^i \,a^j\cr
%\mu\partial_\sigma v &= -\partial_\tau a^i \partial_\sigma a^i
%-{\mu\over 2} \epsilon_{ij} \partial_\sigma a^i \, a^j\,.\cr}}
With the fields $a^i$ 
written in terms of  $X \equiv a^1 + i a^2$ and
$\tilde X \equiv a^1 - i a^2$, \lcact\ becomes
\eqn\lcactx{S_{lc} = \int_\Sigma\, d\tau d\sigma
\,[\,\eta^{\alpha\beta}\, \partial_\alpha X \partial_\beta \tilde X 
- 2 \mu \partial_\tau v - b \mu^2  
- {i\over 2} \mu (\partial_\tau X \,\tilde X - \partial_\tau \tilde X\, X)
+ i \epsilon^{\alpha\beta} \mu \tau \partial_\alpha X \partial_\beta \tilde X
\,]\,.}
The equations of motion and boundary conditions for the
transverse fields are:
\eqn\eomlg{\eqalign{\Box X - i \mu ( \partial_\sigma X - \partial_\tau  X )
&= 0\,,\qquad
\Box \tilde X +
i \mu ( \partial_\sigma \tilde X - \partial_\tau  \tilde X ) = 0\,,\cr
[\, \partial_\sigma X + i\mu\tau \partial_\tau X \,]\,|_{\sigma=0,\pi}
&= 0\,,\qquad
[\,\partial_\sigma \tilde X - i\mu\tau \partial_\tau \tilde X\,]
\,|_{\sigma=0,\pi}
= 0\,,\cr}}
where $\Box\equiv -\partial_\tau^2 + \partial_\sigma^2 =
4 z \bar z \partial_z\partial_{\bar z}\,.$

For large $\tau $ (so that $\tau$ can be considered constant), 
notice the similarity of the boundary condition in \eomlg\
with the boundary condition for an open string in a background
$B$ field. 
Since in the latter case the noncommutativity parameter is proportional to 
the background, this suggests we should expect here 
a noncommutativity parameter which depends on time.

The solution of \eomlg\ is given by the normal mode expansion
for the transverse coordinates $X$ and $\tilde X$, 
to first order in $\mu$:
\eqn\nmex
{\eqalign{X(\sigma,\tau)
&=x_0 + a_0 [ \tau + \mu ( - i\tau\sigma 
+ {i\over 2} \tau^2) ]\cr
&\hskip6pt + \sum_{n\ne0} a_n e^{-in\tau}
[ {i\over n} \cos n\sigma +  \mu ( ( -{1\over 2n^2} - i {\tau\over n} )
\sin n\sigma + ( {i\over 2n^2} + {(\sigma - \tau) \over 2n} ) \cos n\sigma ) ]
+ O(\mu^2)\cr
\tilde X(\sigma,\tau)
&= \tilde x_0 + \tilde a_0 [ \tau - \mu ( - i\tau\sigma 
+ {i\over 2} \tau^2) ]\cr
&\hskip6pt + \sum_{n\ne0} \tilde a_n e^{-in\tau}
[ {i\over n} \cos n\sigma - \mu ( ( -{1\over 2n^2} - i {\tau\over n} )
\sin n\sigma + ( {i\over 2n^2} + {(\sigma - \tau) \over 2n} ) \cos n\sigma ) ]
+ O(\mu^2)\cr}}
We have derived \nmex\ as follows.
In \eomlg\ substitute $X(\sigma,\tau) = e^{i{\mu\over 2} (\tau + \sigma) }\,
\phi(\sigma,\tau)$, and find
\eqn\bcag{ \eqalign {&\Box \phi = 0\,,\cr
&[\,( \partial_\sigma + i\mu\tau \partial_\tau ) \,\phi
+i{\mu\over 2} ( 1 + i\mu \tau ) \,\phi\,] |_{\sigma = 0\,,\pi} = 0\,.\cr}}
One such solution is $\phi(\sigma,\tau) = x_0\,
e^{-i{\mu\over 2} (\tau+\sigma)}$, corresponding to the constant
mode $X(\sigma,\tau) = x_0$.
A general solution to the wave equation $ \Box \phi = 0$ is
\eqn\frew{\phi (\sigma,\tau) = f(\tau + \sigma) + g(\tau - \sigma)\,.}
So the constant solution above corresponds to
$\phi (\sigma,\tau) = f(\tau + \sigma) =
x_0\,e^{-i{\mu\over 2} (\tau+\sigma)}$, and $g(\tau-\sigma) = 0$.
To generate the solutions which provide the
coefficients of $a_0$ and $a_n$ in the normal mode expansion of
$X(\sigma,\tau)$, 
we will try to find solutions $\phi(\sigma,\tau) =
f(\tau + \sigma) +  g(\tau - \sigma)$ satisfying the boundary conditions
\bcag\ via the power series expansions
\eqn\solf{\eqalign{f(\tau + \sigma) &= \sum_{p=0}^\infty C_p
(\tau + \sigma)^p \cr
g(\tau - \sigma) &= \sum_{p=0}^\infty D_p
(\tau - \sigma)^p \cr}}
and
\eqn\solg{\eqalign{f_n(\tau + \sigma) &= e^{-i n (\tau + \sigma)}
\sum_{p=0}^\infty C_p(n) \,(\tau + \sigma)^p \cr
g_n(\tau - \sigma) &= e^{-i n (\tau - \sigma)}\,
\sum_{p=0}^\infty
D_p(n) \,(\tau - \sigma)^p \cr}}
respectively.
A solution of \bcag\ , in the form of \solf\ is
\eqn\newphi{\eqalign{
\mu \phi(\sigma,\tau) &= \mu\tau + \mu^2 [-i {3\over 2} \tau\sigma]
+ \mu^3 \,[ \half \tau^2\sigma + {1\over 6} \sigma^3
- {9\over 8} \tau\sigma^2 - {3\over 8} \tau^3 -{\pi\over 4} (\tau^2
+\sigma^2) \,] \cr
&\hskip3pt + i \mu^4 [ -{1\over 6} \tau^4 + {21\over 16} \tau^3\sigma
-\tau^2\sigma^2 + {21\over 16} \tau\sigma^3   -{1\over 6} \sigma^4
+\pi ( -{3\over 8} \tau^3 + {5\over 8} \tau^2\sigma - {9\over 8} \tau\sigma^2
+ {5\over 24} \sigma^3)\cr
&\hskip25pt  + {\pi^2\over 24} (\tau^2 + \sigma^2)\,]
+ O(\mu^5)\,.\cr}}
where the functions $f$ and $g$ are given by
\eqn\cof{\eqalign{\mu f(\tau) &= {\mu\over 2} \tau
-i{3\over 8} \mu^2\tau^2 -\mu^3 {\pi\over 8}\tau^2
- {5\over 48} \mu^3\tau^3 + i {31\over{3\cdot 128}}\mu^4\tau^4
+ i\mu^4 (-{\pi\over 12} \tau^3 + {\pi^2\over 48}\tau^2 )\,+ O(\mu^5)
\cr
\mu g(\tau) &=  {\mu\over 2} \tau
+ i{3\over 8} \mu^2\tau^2  -\mu^3 {\pi\over 8}\tau^2
- {13\over 48}\mu^3\tau^3 - i {95\over{3\cdot 128}}\mu^4\tau^4
+ i\mu^4 (-{7\pi\over 24} \tau^3 + {\pi^2\over 48}\tau^2)  \,+ O(\mu^5)\,.
\cr}}
These expressions are derived iteratively, by considering the solution
of \bcag\ to some order $\mu^p$, and then integrating the boundary
condition to find the solution to order $\mu^{p+1}$. 
Since finding a general form inarbitrary $p$, and
summing these series to a closed form is difficult, we
work to first order in $\mu$. Note that although $\tau,\sigma$
could be rescaled to essentially eliminate $\mu$, we keep it here to track the
order in the power series solution of \bcag\ . 
The series in \cof\ are reminiscent
of hypergeometric functions.
To derive the coefficient of $a_n$, we use the ansatz \solg\ to find
\eqn\phin{\eqalign{\phi_n(\sigma,\tau) &= i e^{-in\tau}\,
[\cos n\sigma + \mu (( -\tau + {i\over 2n} ) \sin n\sigma
+( -i\sigma + {1\over 2n} ) \cos n\sigma) + O(\mu^2) ]\cr}}
where $\phi_n (\sigma,\tau) = f_n(\tau + \sigma)
+ g_n(\tau -\sigma)$ with
\eqn\fgn{\eqalign
{f_n(\tau) &= i e^{-in\tau}\, [ \half + \mu (-{i\over 2}\tau ) + O(\mu^2) ]\cr
g_n(\tau) &= i e^{-in\tau}\, [ \half + \mu ({i\over 2}\tau + {1\over 2n})
+ O(\mu^2) ]\,.\cr}}
We then construct the normal mode expansion that satisfies
\eomlg\ from 
\eqn\nmexa{\eqalign {X(\sigma,\tau)
&= x _0 + e^{i{\mu\over 2} (\tau + \sigma)}
%\sqrt {2\alpha'}
a_0 \phi(\sigma,\tau)
+ e^{i{\mu\over 2} (\tau + \sigma)}
%i\sqrt {2\alpha'}
\sum_{n\ne0} a_n \phi_n(\sigma,\tau)\,.
\cr}}
{}From \newphi\ and \phin\ , we see that $X(\sigma,\tau)$
is given by an expansion where the coefficients of $a_0$, $a_n$
are themselves a double power series in $\sigma$ and $\tau$.
Although our open string model satisfies an equation of motion
that can be simply related to the one-dimensional wave equation
\eomlg\ , the particular boundary condition
that is required substantially complicates
the form of the solution.
\nmex\ is reproduced by expanding \nmexa\ to first order in $\mu$,
using \newphi\ and \phin\ .
Let $\mu\rightarrow - \mu$ to find $\tilde X(\sigma,\tau)$.

To quantize the theory in standard form, we reinsert the scale
$2 \sqrt {\pi\alpha'}$ so that $X,\tilde X$ become fields with
length dimension, and find the canonical momenta:
\eqn\mom{\eqalign{ P(\sigma,\tau) &= -{\delta S\over \delta \partial_\tau X}
= {1\over 4\pi\alpha'} ( \partial_\tau \tilde X
+ i{\mu\over 2} \tilde X - i \mu \tau \partial_\sigma \tilde X )\cr
\tilde P(\sigma,\tau) &= -{\delta S\over \delta \partial_\tau \tilde X}
= {1\over 4\pi\alpha'} ( \partial_\tau X
- i{\mu\over 2} X + i \mu \tau \partial_\sigma X )\,.\cr}}
To first order in $\mu$,
we can invert the normal mode expansions in \nmex\ as:
\eqn\nm{\eqalign{ ( 1 + {\mu\over 2n} ) \,a_n
&= {1\over {2\pi\sqrt{2\alpha'}}}\,
\int_0^\pi d\sigma\, \cos{n\sigma} \,[ -i n
\,[\, X(\sigma, 0) + X(-\sigma, 0)\,]
+\,[\, 4 \pi\alpha' [ \tilde P (\sigma,0) + \tilde P (-\sigma, 0)] ]]\cr
( 1 - {\mu\over 2n}) \,\tilde a_n &= {1\over {2\pi\sqrt{2\alpha'}}}\,
\int_0^\pi d\sigma\, \cos{n\sigma} \,[ -i n
\,[\,  \tilde X(\sigma, 0) +  \tilde X(-\sigma, 0)\,]
+ \,[\, 4 \pi\alpha' [ P (\sigma, 0 ) + P (-\sigma, 0)] ]]\cr}}
for $n\ne 0$ and
\eqn\nma{\eqalign{x_0 &= {1\over 2\pi} \, \int_0^\pi d\sigma
\,[\, X(\sigma, 0) + X(-\sigma, 0)]\,;\qquad
\tilde x_0 =
{1\over 2\pi} \, \int_0^\pi d\sigma
 \,[\, \tilde X(\sigma, 0) + \tilde X(-\sigma, 0)\,]\cr
{\sqrt{2\alpha'}}\, a_0 - i {\mu\over 2} x_0 &= 2\alpha'  \, 
\int_0^\pi d\sigma
[ \tilde P (\sigma,0) + \tilde P (-\sigma, 0)]
\,;\,\,
{\sqrt{2\alpha'}}\, \tilde a_0 + i {\mu\over 2} \tilde x_0 =
2\alpha'\int_0^\pi d\sigma
[ P (\sigma,0) + P (-\sigma, 0)]\,.\cr
}}

The commutation relations which follow from canonical
quantization 
$[ X(\sigma,\tau), P(\sigma',\tau) ] = i \delta (\sigma-\sigma'),\,
[\tilde X(\sigma,\tau), \tilde P(\sigma',\tau) ] = i \delta (\sigma-\sigma')$
are:
\eqn\commu{\eqalign{[a_m, \tilde a_n ]
&= 2\, (m - \mu) \delta_{m,-n} \,;\,\qquad
[a_m, a_n ] =  [\tilde a_m, \tilde a_n ] = 0\,;\cr
[x_0,\tilde x_0 ] &=  0 \,;\qquad
[a_n, x_0] = [a_n, \tilde x_0] = [\tilde a_n, x_0] = [\tilde a_n, \tilde x_0]
=0 \, {\rm for }\, n\ne 0\,;\cr
[x_0, \tilde a_0] &= i 2 \sqrt{2\alpha'} = [\tilde x_0, a_0]\,;\qquad
[x_0, a_0] = [\tilde x_0, \tilde a_0] = 0\,.\cr}}

\newsec{The Propagator on the Disk}

Having found a mode expansion, we compute the propagator,
along the lines of  {\refs{\dn}}.
In $z,\bar z$ coordinates (where $z$ is in the upper half plane,
since $0\le\sigma\le\pi$),
the equation of motion and boundary conditions
for the propagator are:
\eqn\eomp{\eqalign{&4 z \bar z \partial_z  \partial_{\bar z} X
- 2 \mu \bar z \,\partial_{\bar z} X = 0\,,\qquad
4 z \bar z \partial_z  \partial_{\bar z}
\tilde X + 2 \mu \bar z \,\partial_{\bar z} \tilde X = 0\cr
&( \partial_z - \partial_{\bar z}) X + {\mu\over 2} \ln {z\bar z}
\,(\partial_z + \partial_{\bar z}) X |_{z=\bar z} = 0\,,\qquad
( \partial_z - \partial_{\bar z}) \tilde X - {\mu\over 2} \ln {z\bar z}
\,(\partial_z + \partial_{\bar z}) \tilde X |_{z=\bar z} = 0\cr
& 4 \partial_z  \partial_{\bar z}
\, < X (z,\bar z) \tilde X (\zeta,\bar\zeta) >
- 2 \,\mu z^{-1} \,\partial_{\bar z}  < X (z,\bar z) \tilde X (\zeta,\bar\zeta) >
= -2\pi \alpha'
\delta^2 (z-\zeta)\cr
&[\,( \partial_z - \partial_{\bar z}) \,
<X (z,\bar z) \tilde X (\zeta,\bar\zeta) >
+ {\mu\over 2} \ln {z\bar z} \,(\partial_z + \partial_{\bar z})\,
<X (z,\bar z) \tilde X (\zeta,\bar\zeta) >]\,
|_{z=\bar z} = 0\,.\cr}}
We will compute the propagator on the disk, and will
use $z=e^{i(\tau + \sigma)}$, $\bar z=e^{i(\tau - \sigma)}$,
$\zeta=e^{i(\tau' + \sigma')}$ and $\bar\zeta=e^{i(\tau' - \sigma')}$.
In the above boundary conditions,
the notation $|_{z=\bar z}$ denotes $ z = |z|, \bar z = |z|$ at the
$\sigma = 0$ endpoint and $z = |z| e^{i\pi}, \bar z = |z|e^{-i\pi} $ at
$\sigma = \pi$.
Assuming the commutation relations in \commu\ ,
then for $|z|> |\zeta|$, the propagator to order $\mu$ is
\eqn\propa{\eqalign{< X (z,\bar z) \tilde X (\zeta,\bar\zeta) >
&= {\sqrt{2\alpha'}}\,[a_0,\tilde x_0] (\tau + \mu ( -i \tau\sigma
+ {i\over 2}\tau^2) )\cr
&\hskip4pt + 2\alpha' \,\sum_{n=1}^\infty  [a_n,\tilde a_{m}]
\, e^{-i n \tau} e^{-i m \tau')}\,
[\,-{1\over n m} \cos n\sigma \cos m\sigma'\cr
&\hskip15pt + i{\mu\over m} \cos m\sigma' (\, ( -{1\over 2n^2}
- {i\tau\over n} ) \sin n\sigma
+ ( {i\over 2n^2} + {(\sigma - \tau)\over 2n} )\cos n\sigma \,)\cr
&\hskip15pt - i{\mu\over n} \cos n\sigma (\, ( -{1\over 2m^2}
- {i\tau'\over m} ) \sin m\sigma'
+ ( {i\over 2m^2} + {(\sigma' - \tau')\over 2m}) \cos m\sigma' \,) \,]\cr
&\hskip4pt + \mu ( c_1\tau + c_0 ) \cr
&= - i 4\alpha' \,
(\tau + \mu ( -i \tau\sigma + {i\over 2}\tau^2) )\cr
&\hskip4pt + 4\alpha' \sum_{n=1}^\infty  e^{-i n (\tau - \tau')}
[\,{1\over n} \cos n\sigma \cos n\sigma'\cr
&\hskip15pt + i\mu \cos n\sigma' (\, ( {1\over 2n^2}
+ {i\tau\over n} ) \sin n\sigma
- ( {i\over 2n^2} + {(\sigma - \tau)\over 2n} )\cos n\sigma \,)\cr
&\hskip15pt - i\mu \cos n\sigma (\, ( {1\over 2n^2}
- {i\tau'\over n} ) \sin n\sigma'
+ ( {i\over 2n^2} - {(\sigma' - \tau')\over 2n}) \cos n\sigma' \,) \cr
&\hskip15pt -{\mu\over n^2} \cos n\sigma \cos n\sigma'\,]\cr
&\hskip4pt + \mu ( c_1\tau + c_0 )\,. \cr
}}

We are free to add the function $ \mu (c_1 \tau + c_0 ) $ to the expression
since it does not affect the equation of motion or the boundary
condition for the propagator to first order in $\mu$.
For  $|z|> |\zeta|$, the expression for
$< \tilde X (z,\bar z) X (\zeta,\bar\zeta) >$ is given by
letting $\mu\rightarrow -\mu$ in the above propagator.
In the $\mu\rightarrow 0$ limit, these propagators reduce to the
open bosonic string propagator
$\lim_{\mu\rightarrow 0} < X (z,\bar z) \tilde X (\zeta,\bar\zeta) >
= -2\alpha' ( \,\ln |z-\zeta| + \ln |z-\bar\zeta| \,)$.

\newsec{Time-Dependent Noncommutativity}

To evaluate the noncommutativity parameter as defined from
time ordering \refs{\sch,\sw},
we consider the propagator on the worldsheet boundary
at $\sigma = 0$, then $z = |z| = e^{i\tau}
\equiv {\cal T}$,
and $\zeta =  e^{i(\tau' +\sigma')} =
|\zeta | = e^{i\tau'} = {\cal T'}$, so ${\cal T}, {\cal T'}>0$.
We will also consider the propagator at
$\sigma = \pi$, then $z = |z| e^{i\pi}= {\cal T}$ and
$\zeta = |\zeta |  e^{i\pi} = {\cal T'}$ so here ${\cal T},{\cal T'} <0$.
Note that ${\cal T}$ is different from the worldsheet time
$\tau$.
\eqn\probn
{\eqalign{< X (z,\bar z) \tilde X (\zeta,\bar\zeta) >|_{\sigma = 0}
&= -i4\alpha'\, ( \tau + \mu {i\over 2}\tau^2)  +  \mu ( c_1\tau + c_0)\cr
&\hskip6pt - 4\alpha' \ln ( 1 - e^{- i (\tau - \tau')} )
- 2\alpha' \mu
\,i (\tau - \tau')\, \ln ( 1 - e^{-i (\tau - \tau')})\cr
&= - 4 \alpha' \ln ({\cal T} - {\cal T'})\cr
&\hskip6pt + \mu ( - 2 \alpha' \ln^2 {\cal T}
- 2 \alpha'
\ln ({{\cal T}\over {\cal T'}}) \, \ln ( 1 - {{\cal T'}\over {\cal T}})
\, + ( -c_1 i \ln {\cal T} + c_0) )\cr
< \tilde X (z,\bar z) X (\zeta,\bar\zeta) >|_{\sigma = 0}
&=  -i4\alpha'\, ( \tau - \mu {i\over 2}\tau^2) - \mu ( c_1\tau + c_0)\cr
&\hskip6pt - 4\alpha' \ln ( 1 - e^{- i (\tau - \tau')} )
+ 2\alpha' \mu
\,i (\tau - \tau') \, \ln ( 1 - e^{-i (\tau - \tau')})\cr
}}
Then at $\sigma = 0$:
\eqn\scho{\eqalign{[ X ({\cal T}) ,  \tilde X ({\cal T})]
&= T (  X ({\cal T}) \, \tilde X ({\cal T}^-) -  X ({\cal T})
\tilde X ({\cal T}^+) )\cr
&\equiv \lim_{\epsilon\rightarrow 0}
(  < X ({\cal T}) \, \tilde X ({\cal T}-\epsilon ) > -
< \tilde X ({\cal T} + \epsilon )\, X ({\cal T}) > )
\, , \quad ({\rm for}\, \epsilon > 0)\cr
&= \mu \,
(-4 i \alpha' ) ( \pi \ln {\cal T} - i\ln^2{\cal T})
\cr
&= \mu \,  4\alpha' \,( \pi \tau +  \tau^2)
\equiv \Theta\,,\cr}}
where we chose $c_1 = 2\pi\alpha'$, $c_0 = 0$,
and use $\lim_{\epsilon\rightarrow 0} (\ln ( 1 + \epsilon)
\ln \epsilon ) = 0$.
The noncommutativity parameter $\Theta$ is time-dependent.

At  $\sigma = \pi$:
\eqn\probnp
{\eqalign{< X (z,\bar z) \tilde X (\zeta,\bar\zeta) >|_{\sigma = \pi}
&= -i4\alpha'\, ( \tau + \mu ( - i\tau\pi
+ {i\over 2}\tau^2) ) + \mu (c_1\tau + c_0) \cr
&\hskip6pt - 4\alpha' \ln ( 1 - e^{i (\tau' - \tau)} )\cr
&\hskip6pt - 2\alpha' \mu
\,i (\tau - \tau')\,  \ln ( 1 - e^{-i (\tau - \tau')})\cr
< \tilde X (z,\bar z) X (\zeta,\bar\zeta) >|_{\sigma = \pi}
&=  -i4\alpha'\, ( \tau - \mu ( - i\tau\pi
+ {i\over 2}\tau^2) ) + \mu (c_1\tau + c_0)\cr
&\hskip6pt - 4\alpha' \ln ( 1 - e^{i (\tau' - \tau)} )\cr
&\hskip6pt + 2\alpha' \mu
\,i (\tau - \tau')\, \ln ( 1 - e^{-i (\tau - \tau')})\cr
}}
\eqn\schpi{\eqalign{[ X ({\cal T}) ,  \tilde X ({\cal T})]
&= T (  X ({\cal T}) \, \tilde X ({\cal T}^-) -  X ({\cal T})
\tilde X ({\cal T}^+) )\cr
&\equiv \lim_{\epsilon\rightarrow 0}
(  < X ({\cal T}) \, \tilde X ({\cal T}+\epsilon ) > -
< \tilde X ({\cal T} - \epsilon )\, X ({\cal T}) > )
\, , \quad ({\rm for}\, \epsilon > 0)\cr
&= (-i 4\alpha')\, \mu [\, -\pi \ln{\cal T}
- i \ln^2{\cal T} \,]\cr
&= \mu \,\, 4\alpha' \, (\,-\pi \tau + \tau^2)\,
\cr}}
Thus for small $\mu$, we have:
\eqn\ncpmu{\eqalign{\Theta&=  \mu \,\, 4\alpha' \,( \pi \tau +  \tau^2)
\,\qquad {\rm at}\, \sigma = 0\,,\cr
\Theta&= \mu \,\, 4\alpha' \, (\,-\pi \tau + \tau^2)\,
\quad {\rm at}\, \sigma = \pi\,.\cr}}
For small $\tau$, the theta parameter at the
$\sigma = 0$ end of the string is minus that at the $\sigma = \pi$ end.
This is the case for the neutral string in a constant background
$B$ field as well. In fact, although we have worked only to
lowest order in $\mu$, we can see directly from the equations of
motion and boundary conditions (in $z$, $\bar z$) variables in \eomp\ ,
that in the limit of large $z$, {\it i.e.}
large $i\tau$, a limit for which $z^{-1}\rightarrow\ 0$, that
the system reduces to the neutral string with
the identification $- \mu \tau = B$, a constant.
(In the large $\tau$ limit, we note that $\ln |z|$ is approximately
constant, in the sense that it is changing slowly, {\it i.e.} its derivative
$ |z|^{-1}$ is small. Therefore, for large $\tau$ the noncommutativity
parameter becomes constant, and our model is similar to the
neutral string.)
For large $\tau$, using the neutral string expressions, 
we find the noncommutativity parameter 
be time-dependent:
\eqn\ncp{\eqalign{\Theta&= -4\alpha' \pi B = 4\alpha' \mu \pi \tau
\,\,\,\quad {\rm at} \,\,\,\sigma = 0\,,\cr
\Theta&= 4\alpha' \pi B = - 4\alpha' \mu \pi \tau
\,\,\,\quad{\rm at} \,\,\sigma = \pi\,.\cr}}

We have shown that our model 
exhibits noncommutativity for both small and large $\tau$.
The expectation is that the model will remain noncommutative 
with a time-dependent noncommutativity parameter for all times.  

\vskip8pt

%NOTE (I'm not suggesting we include this, but we need to keep it in mind.):
%We can check that since \newphi\ implies
%\eqn\newder{\eqalign{\partial_\sigma\phi &= \mu^2 [-i{3\over 2}\tau]
%+ \mu^3 [ \half \tau^2 + \half\sigma^2 -{9\over 4} \tau\sigma
%- {\pi\over 2}\sigma \,]\cr
%&\hskip3pt  + i \mu^4 [{21\over 16} \tau^3
%- 2\tau^2\sigma + {3\cdot 21\over 16} \tau\sigma^2   -{2\over 3} \sigma^3
%+ \pi ( {5\over 8} \tau^2 - {9\over 4} \tau\sigma
%+ {5\over 8} \sigma^2)
%+ {\pi^2\over 12} \sigma\,] \cr
%&\hskip3pt + \mu^5[]\cr
%\partial_\tau\phi &= \mu + \mu^2 [-i{3\over 2}\sigma] + \mu^3
%[ \tau\sigma - {9\over 8} \sigma^2 -{9\over 8} \tau^2 - {\pi\over 2}
%\tau ] + \mu^4 [ ]\cr
%i\mu\tau\partial_\tau\phi &= \mu^2[i\tau]  + \mu^3 [{3\over 2}\tau\sigma]
%+ i\mu^4
%[ \tau^2\sigma - {9\over 8} \tau\sigma^2-{9\over 8} \tau^3 - {\pi\over 2}
%\tau^2 ] + \mu^5 [ ]\cr
%i{\mu\over 2}\phi &= \mu^2 [{i\over 2}\tau ] + \mu^3 [{3\over 4} \tau\sigma]
%+ i \mu^4 [ {1\over 4} \tau^2\sigma + {1\over 12} \sigma^3
%- {9\over 16} \tau\sigma^2 - {3\over 16} \tau^3 -{\pi\over 8} (\tau^2
%+\sigma^2) \,]  + \mu^5 [ ]\cr
%-{\mu^2\over 2}\tau\phi &= \mu^3 [{-\half}\tau^2] + \mu^4
%[i{3\over 4}\tau^2\sigma] + \mu^5 [ ]\cr}}
%then \newphi\ satisfies \bcag\ to order $\mu^4$.

\vskip30pt

\leftline{\bf Acknowledgements}
It is  a pleasure to thank Edward Witten for discussions.
LD is grateful to Princeton University and the Institute for Advanced Study
for their hospitality during the summer 2002, and to the Aspen Center for
Physics.  She was supported in part
by U.S. Department of Energy, Grant No. DE-FG 05-85ER40219/Task A.
CRN is supported in part by NSF grant PHY-0140311. 
[Any opinion, findings, and conclusions or recommendations expressed in this 
material are those of the authors and do not necessarily reflect the views 
of the National Science Foundation.] 
\listrefs

\bye